\documentclass[aps,prl,reprint,groupedaddress]{revtex4-1}

\usepackage[pdftex]{graphicx}
\usepackage{graphics}
\usepackage{ upgreek }
\usepackage{ textcomp }
\usepackage{ tabularx }

\begin{document}

\title{One-dimensional helical transport in topological insulator nanowire interferometers}

\author{Seung Sae Hong,$^1$ Yi Zhang,$^2$ Judy J. Cha,$^3$ Xiao-Liang Qi,$^2$ and Yi Cui$^3$  }
\affiliation{ $^1$Department of Applied Physics, Stanford University, Stanford, California 94305, USA}
\affiliation{ $^2$Department of Physics, Stanford University, Stanford, California 94305, USA}
\affiliation{ $^3$Department of Materials Science and Engineering, Stanford University, Stanford, California 94305, USA}

\date{\today}

\begin{abstract}

The discovery of three-dimensional (3D) topological insulators opens a gateway to generate unusual phases and particles made of the helical surface electrons, proposing new applications using unusual spin nature. Demonstration of the helical electron transport is a crucial step to both physics and device applications of topological insulators. Topological insulator nanowires, of which spin-textured surface electrons form 1D band manipulated by enclosed magnetic flux, offer a unique nanoscale platform to realize quantum transport of spin-momentum locking nature. Here, we report an observation of a topologically protected 1D mode of surface electrons in topological insulator nanowires existing at only two values of half magnetic quantum flux ($\pm$\textit{h/}2\textit{e}) due to a spin Berry's phase ($\pi$). The helical 1D mode is robust against disorder but fragile against a perpendicular magnetic field breaking time-reversal-symmetry. This result demonstrates a device with robust and easily accessible 1D helical electronic states from 3D topological insulators, a unique nanoscale electronic system to study topological phenomena.

\end{abstract}

\maketitle

On the surface of topological insulators (TIs), electron spin is locked perpendicular to momentum, resulting in the suppression of electron backscattering from nonmagnetic impurities\cite{Moore2010, Hasan2010, Qi2010, Konig2007, Hsieh2008}. Such spin-textured surface states have been studied extensively by surface probing techniques such as angle-resolved photoemission spectroscopy (ARPES) and scanning tunneling microscopy (STM)\cite{Hsieh2008, Hsieh2009, Chen2009, Chen2010, Roushan2009, Zhang2009PRL, Alpichshev20010, McIver2012}. Electronic device demonstrating topologically distinctive properties, crucial to both physics and device applications, however, has lagged behind in part due to the dominance of large bulk carrier concentration. Recent studies on Shubnikov-de Haas (SdH)\cite{Analytis2010, Qu2010, Sacepe2011, Taskin2012} and Aharonov-Bohm (AB)\cite{Peng2009, Xiu2011} oscillations have provided evidence of surface electron conduction although the unique spin nature remains undemonstrated in transport. 

	Aharonov-Bohm (AB) oscillations\cite{Washburn1986, Aronov1987, Bachtold1999} are particularly interesting quantum phenomena to manifest the topological properties in TI nanowire devices. In a closed trajectory encircling certain magnetic flux, coherent electrons interfere by themselves and their phase is periodically modulated by the encircled magnetic flux. The signature of AB effect is magnetoconductance oscillation of flux quantum (\textit{\textit{h/e}}) periodicity, observed first in 2D metal rings\cite{Washburn1986}. Another well-known example is a metallic cylinder that is regarded as a sum of infinite number of 2D metal rings\cite{Aronov1987}. In a cylindrical conductor, \textit{h/}2\textit{e} period oscillations, of weak (anti)localization effect from clockwise/counterclockwise trajectories, are the dominant interferences known as Altshuler-Aronov-Spivak (AAS) oscillations. \textit{\textit{h/e}} period oscillations are usually suppressed in cylindrical conductors as different slices of the metal cylinder (effectively 2D metal ring) generate \textit{\textit{h/e}} period oscillations of random phases, canceling each other\cite{Aronov1987, Bachtold1999}. Unexpectedly, the first experimental observation of AB effect in TI nanowires reported \textit{\textit{h/e}} periodicity in magneto-oscillations\cite{Peng2009, Xiu2011}. This observation has initiated a number of theoretical studies\cite{Bardarson2010, Zhang2010PRL, Rosenberg2010, Egger2010}, and it is believed that the \textit{\textit{h/e}} period oscillations is indeed closely related to the 1D band formation of the surface electrons.

The characteristic band structure of the topological surface states is conical band dispersion in 2D\cite{Zhang2009nphy, Xia2009} (Dirac cone, Fig. 1a), but the band structure experiences significant modification when the surface electrons are confined in 1D nanostructures of TI. 2D Dirac Fermions on the TI nanowire surface form discrete 1D subbands due to the quantum-confined periodic boundary condition along the perimeter direction, when the trajectory is short enough to maintain ballistic condition\cite{Bardarson2010, Zhang2010PRL, Rosenberg2010, Egger2010} (Fig. 1b). This is exactly the same case as the transition from graphene to a carbon nanotube\cite{Dekker1999}, except the fact that TI surface is spin-textured (helical). The spin-momentum locking property of the TI surface electrons adds a spin Berry's phase ($\pi$) to the electron wave function, by the 2$\pi$ rotation of spin along the nanowire perimeter. Counting the additional phase (2$\pi$$\Phi$/$\Phi$$_0$, $\Phi$$_0$=\textit{\textit{h/e}}) from the magnetic flux ($\Phi$) enclosed by the surface electron trajectory (nanowire perimeter \textit{L}), topological surface states in nanowires have the 1D band dispersion as:

\begin{equation}
E (n, k, \Phi) = \pm hv_F \sqrt[]{ (\frac{k}{2\pi})^2 + (\frac{n+1/2-\Phi/\Phi_0}{L})^2 } ,
\end{equation}

where \textit{n} is an integer, the half integer (1/2) term is from the spin Berry's phase, and \textit{h}, v$_F$ are the plank constant and Fermi velocity respectively.

\begin{figure*}
 \includegraphics[width=1\linewidth]{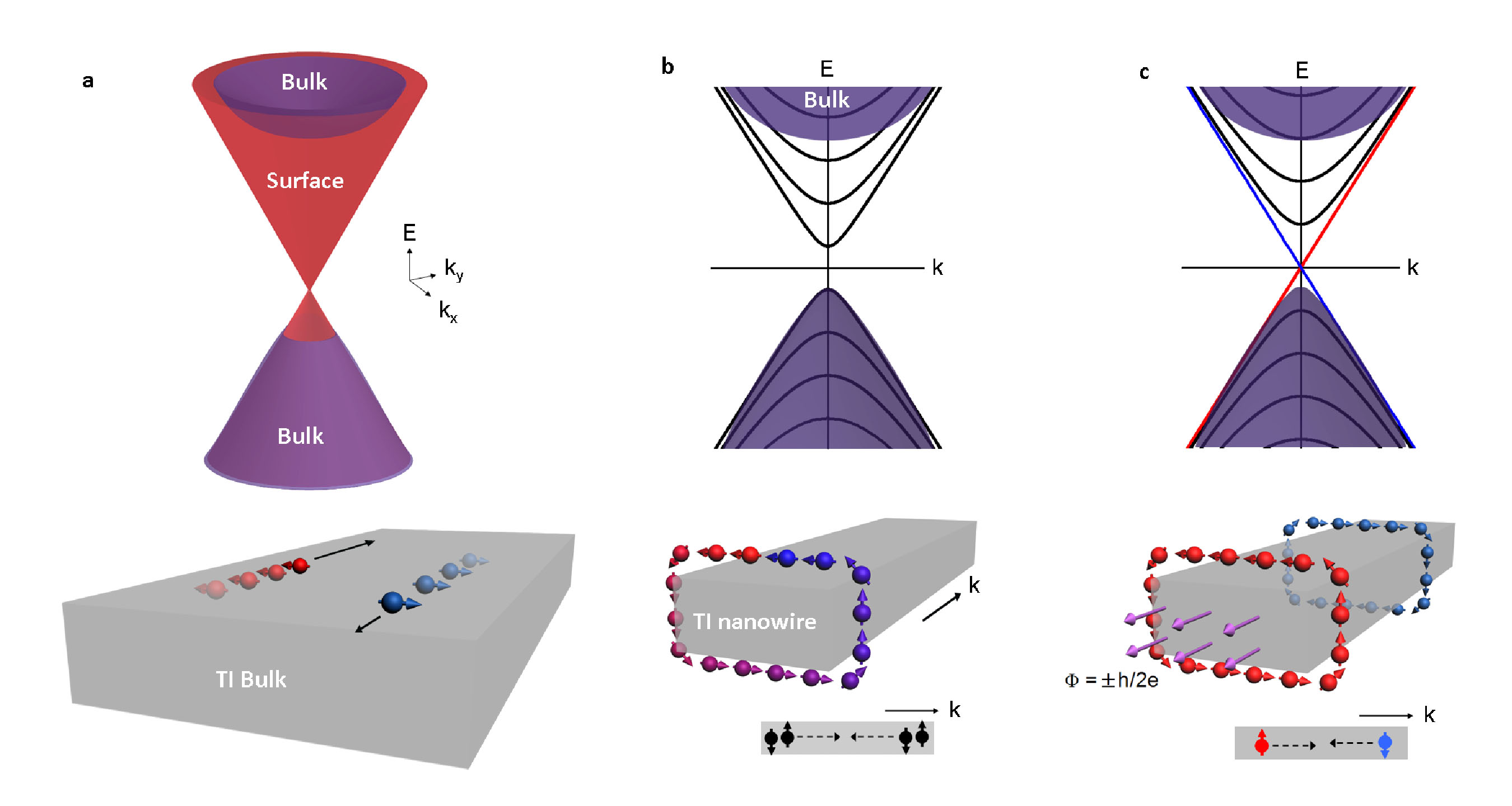}
 \caption{Topological surface band evolution in TI nanowires (\textbf{a}) Top: schematic band structure of bulk topological insulator (Bi$_2$Se$_3$). Surface states (red) exist in parallel with bulk conduction/valence band (purple). Bottom: electrons in the surface band are moving in two-dimensional surface, of which spin is defined by electron momentum (no spin degeneracy).  (\textbf{b}) Top: band structure of a topological insulator nanowire (with no magnetic field). Topological surface band transforms to discrete 1D subbands (black color bands) with spin degeneracy. Bottom: electron spin is constrained in the tangent plane picking up a $\pi$ Berry's phase by a 2$\pi$ rotation of electrons along the perimeter, which opens a gap in the 1D modes. (\textbf{c}) Top: band structure of a topological insulator nanowire (with magnetic flux $\Phi$ = $\pm$\textit{h/}2\textit{e}). The gapless bands (red and blue) is not spin degenerated and topologically nontrivial, referred as 1D helical mode. Bottom: electrons of opposite spin orientations propagate in an opposite manner.}
 \label{fig1}
\end{figure*}

The 1D band structure of the surface states (Eq. 1) explains uncommon \textit{\textit{h/e}} period oscillations in TI nanowire and hints an exotic electronic state of topological nature. The 1D bands and their density of states (DOS) periodically changes by magnetic flux, resulting in magneto-oscillations of $\Phi$$_0$=\textit{\textit{h/e}} period as reported previously\cite{Peng2009, Xiu2011}. These oscillations are mainly due to the parabolic 1D bands and cannot be distinguishable from AB effects from ordinary (not topological) electrons. However, at a specific value of magnetic flux ($\Phi$=$\pm$\textit{h/}2\textit{e}, $\pm$3\textit{h/}2\textit{e}, …), the AB phase shift induced by the magnetic flux ($\pi$) cancels the spin Berry's phase ($\pi$) resulting in a single, gapless 1D mode (Fig. 1c). This gapless 1D mode is predicted to be topologically nontrivial electronic state of no spin-degeneracy\cite{Bardarson2010, Zhang2010PRL, Rosenberg2010, Egger2010}. The helical 1D mode at $\Phi$=$\pm$\textit{h}/2\textit{e}, the key in manifesting helical nature of the surface electrons and topological properties via transport, can be detected at two different situations. First, we expect to see pure helical 1D transport when the Fermi level of nanowire is close to the Dirac point, as the Fermi level only crosses the helical gapless 1D band. The other case is that the nanowire contains enough (elastic) disorders. Strong disorders in materials would suppress quantum interferences from parabolic (spin-degenerated) bands. However, the helical 1D mode is robust against disorder due to the topological protection, and should be detectable when all other 1D bands are localized\cite{Bardarson2010, Zhang2010PRL}. In both cases, we expect to observe conductance peaks emerging at the two values of half quantum flux ($\Phi$=$\pm$\textit{h}/2\textit{e}).

In this letter, we provide three main experimental observations to prove the topological nature of surface electrons via transport in nanowires. First, in a TI nanowire, of which Fermi level approaches to the Dirac point, we observe large conductance peaks at $\Phi$=$\pm$\textit{h}/2\textit{e}, originated from 1D helical mode. Second, we induce stronger disorder by sample aging. In nanowire devices of strong disorders, the helical 1D mode peaks are detected at $\Phi$=$\pm$\textit{h}/2\textit{e} while both \textit{\textit{h/e}} period oscillations (from topologically-trivial 1D modes) and \textit{h/}2\textit{e} period oscillations (AAS oscillations) are suppressed by disorder. Finally, by applying perpendicular magnetic fields, we find that the helical 1D mode peaks from disordered nanowires are fragile to TRS breaking, meaning topological nature directly related to TRS. These results demonstrate a nanowire quantum device manifesting topologically unique nature.

\begin{figure*}
 \includegraphics[width=0.8\linewidth]{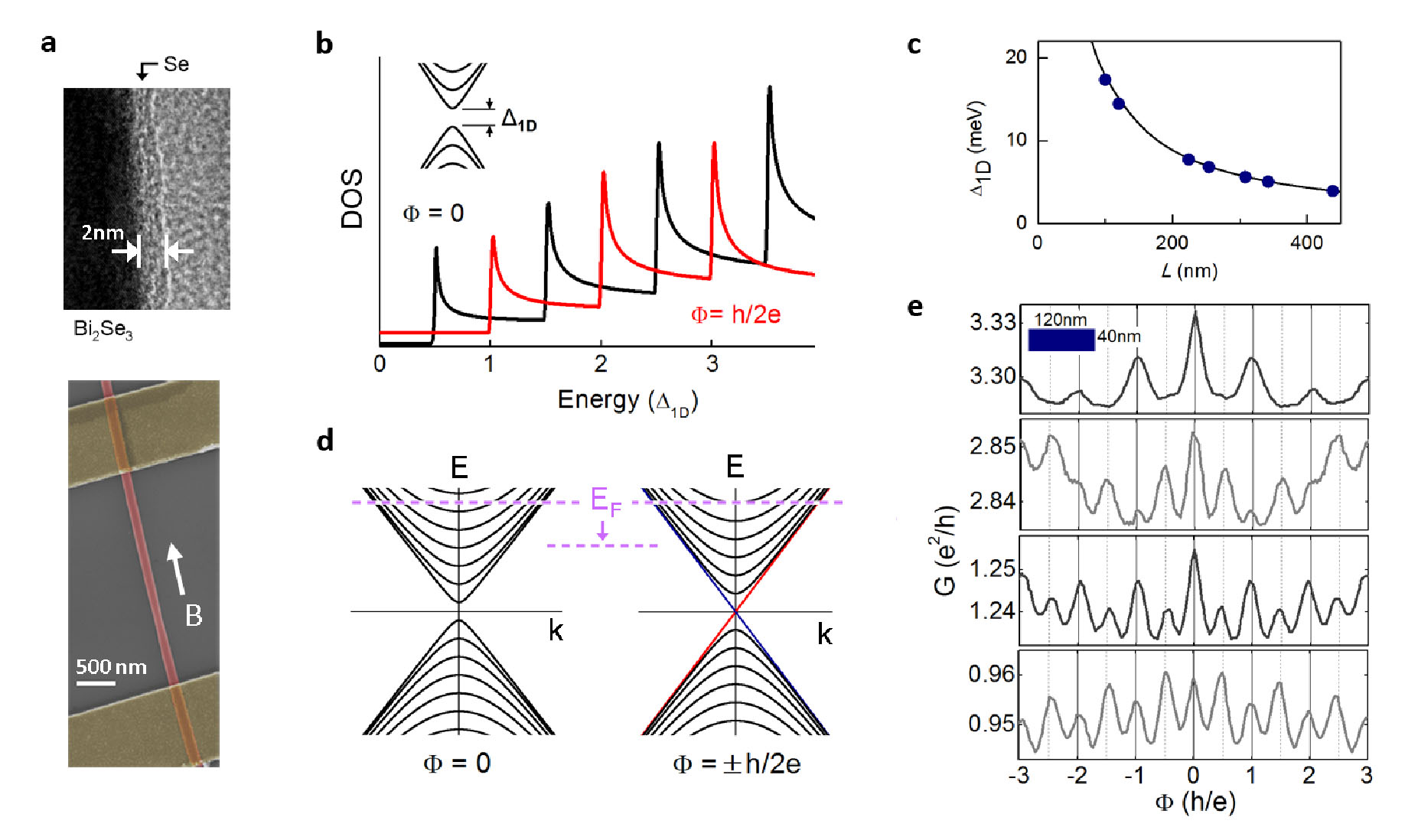}
 \caption{Magnetoconductance from TI nanowire 1D modes of surface electrons: far above the Dirac point (\textbf{a}) Top: TEM image of Bi$_2$Se$_3$ core - Se shell nanowire. Bottom: a representative image of devices (false-colored scanning electron microscope image). (\textbf{b}) Density of states (DOS) of the surface electron 1D mode, as a function of 1D band gap at zero flux($\Delta$$_{1D}$). (\textbf{c}) 1D bandgap ($\Delta$$_{1D}$) as a function of nanowire perimeter (\textit{L}). Blue dots are the estimated energy gaps of seven nanowire devices . (\textbf{d}) schematic band structure of 1D modes in TI nanowires of larger perimeter (\textit{L} $\sim$ 300 nm) with two quantum flux value (0 and $\pm$\textit{h/}2\textit{e}). Fermi level (E$_F$) is crossing many subbands, and the location of E$_F$ is tuned by gating. (\textbf{e}) Magnetoconductance oscillations at four different gate voltages (device A, V$_G$=-40V, -45V, -85V, -95V). Inset: the cross section of the nanowire (120 nm $\times$ 40nm).}
 \label{fig2}
\end{figure*}

Two experimental conditions, a) ballistic transport along the nanowire perimeter and b) Fermi level located at the Dirac point, should be satisfied to realize helical transport proposed by the theoretical predictions\cite{Bardarson2010}. To enhance the surface electron mobility and satisfy the ballistic condition, we synthesize heterostructure nanowires consisting of a Bi$_2$Se$_3$\cite{Zhang2009nphy, Xia2009} core and an amorphous Se shell. A 2 nm-thick Se layer is coated \textit{in-situ} after an initial vapor-liquid-solid growth of a Bi$_2$Se$_3$ nanowire\cite{Kong2010}, keeping a clean Bi$_2$Se$_3$ surface (Fig. 2a). High surface mobility is confirmed from the core - shell nanowires (Supporting information), leading long mean free path surpassing the nanowire perimeter. We emphasize here that the nanowires are not ballistic along the nanowire length, but the ballistic condition along the nanowire perimeter is sufficient enough to observe 1D band structure and relevant topological physics.

The control of Fermi level is another issue in experiment, as Bi$_2$Se$_3$ nanomaterials are always n-type conductors. For an effective tuning of the Fermi level by SiO$_2$ gating, we selectively choose the dimension of nanowire devices. First, the length of nanowires should be long enough ($>$ 1 $\mu$m) to fully deplete the nanowires by field effect. The width and thickness of nanowires are very critical as well. The device Fermi level can be tunable by back gating to reach the Dirac point\cite{Sacepe2011, Hong2012, Kim2012}, but the electrostatic gating usually results in a slight difference of top and bottom surfaces' Fermi level\cite{Sacepe2011, Hong2012}. To study the helical 1D mode at the Dirac point, it is necessary to have 1D band gap ($\Delta$$_{1D}$) large enough to locate entire nanowire's Fermi level (of top and bottom surfaces) within the bandgap. The 1D band gap ($\Delta$$_{1D}$ = hv$_F$/\textit{L}) is determined by the nanowire perimeter (\textit{L} = 100 - 450 nm), as shown in Fig. 2b-c, ranging from 17 meV to 4 meV. We observe clean signature of helical 1D mode from devices of the largest gap energy (15 - 17 meV) which will be discussed in the next paragraphs.

We first measured electron transport of TI nanowire devices of similar sizes to the devices from previous studies\cite{Peng2009}, with 1D band gaps of 6 meV (40 nm height x 120 nm width) (Fig. 2d). In magnetic field along the nanowire length, both \textit{h/}2\textit{e} period and \textit{\textit{h/e}} period oscillations are observed (Fig. 2e).  Oscillations of both periods are anticipated by theory\cite{Bardarson2010, Zhang2010PRL} - \textit{h/}2\textit{e} period from AAS effect and \textit{\textit{h/e}} period from periodic modulation of 1D subband density of state (DOS) by magnetic flux (Fig. 2b). \textit{\textit{h/e}} period oscillations, which are also reported by the previous works\cite{Peng2009, Xiu2011}, show small amplitude, $\sim$ 1 $\%$ of total conductance. The small oscillation amplitude is due to two factors: a) bulk electron contribution and b) density of state effect from many 1D subbands. As the Fermi level crosses multiple 1D subbands (Fig. 2d), the change of DOS induced by the magnetic flux ($\sim$ one subband's DOS) is much smaller than the total DOS (entire subbands' DOS), resulting in small oscillation amplitude. The DOS effect in \textit{h/e} period oscillations becomes more evident when the Fermi level is tuned by back gate (Fig. 2e). The maxima location of \textit{h/e} period oscillations switches discretely between integer quantum flux ($\Phi$ = 0, $\pm$\textit{\textit{h/e}}, $\pm$2\textit{\textit{h/e}}, …) and half quantum flux ($\Phi$ = $\pm$\textit{h/}2\textit{e}, $\pm$3\textit{h/}2\textit{e}, …). The \textit{h/e} period oscillations peaking at either integer quantum flux or half quantum flux is the signature of 1D band effect made of surface electrons, when the Fermi level is far above from the Dirac point. This experimental results are explicitly predicted by the 1D DOS theory of TI nanowires\cite{Bardarson2010} (Fig. 2b).

Then we focus on electron transport of TI nanowires at the Dirac point. In a TI nanowire of which Fermi level is at the Dirac point, there exists no 1D electronic state under zero magnetic field (Fig. 2b). But the gapped 1D band starts to be modified by adding magnetic flux, and gapless (helical) 1D state is formed at $\Phi$ = $\pm$\textit{h/}2\textit{e}(Fig. 1c). Thus magnetoconductance of TI nanowires at the Dirac point will have conductance peaks only peaking at $\Phi$ = $\pm$\textit{h/}2\textit{e}\cite{Bardarson2010}. The magnitude of the peak is expected to be large, because only one 1D mode contributes to the magnetoconductance.

\begin{figure}
\centering
\includegraphics[width=88mm]{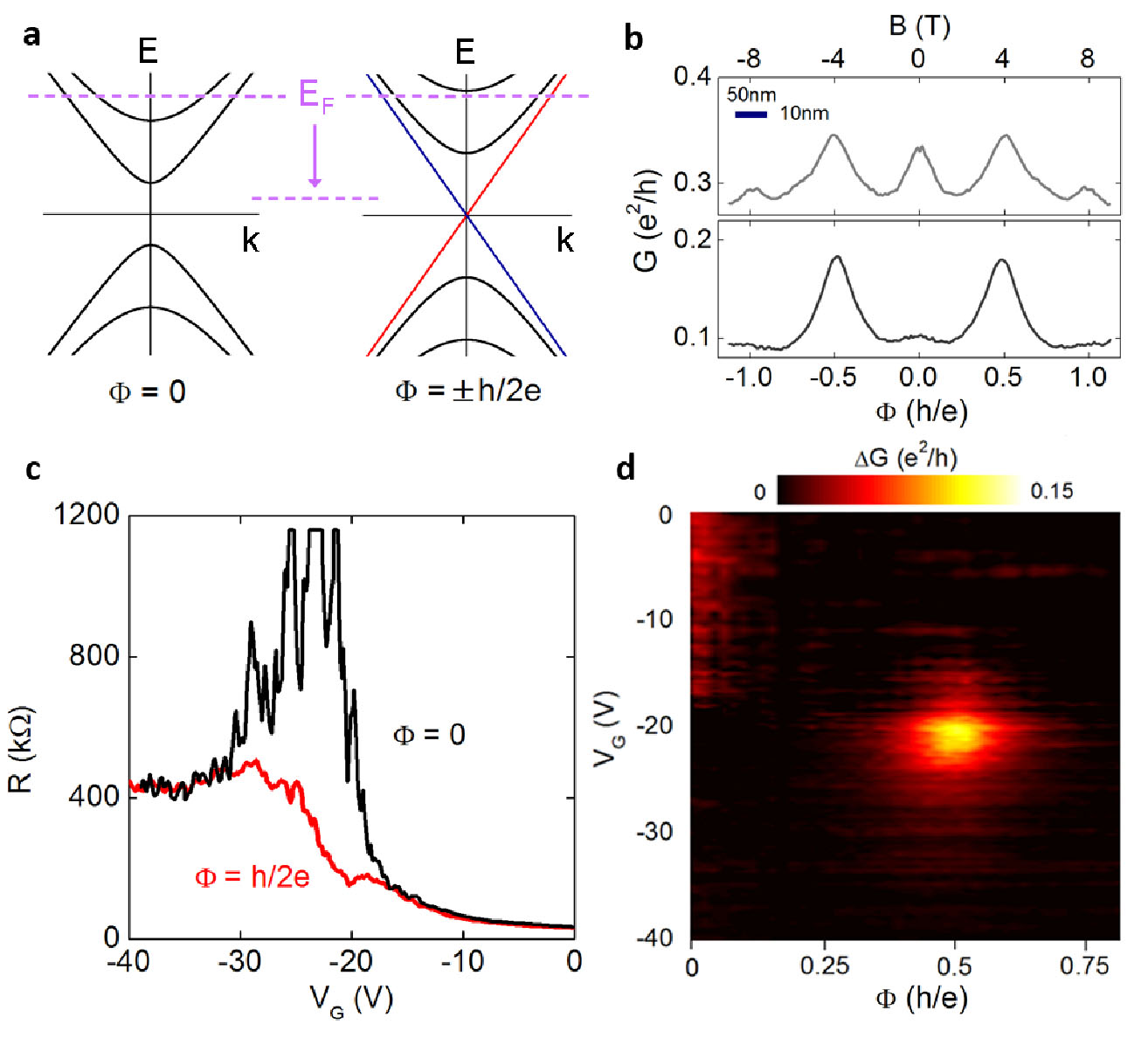}

\caption{Magnetoconductance from TI nanowire 1D modes of surface electrons: at the Dirac point (\textbf{a}) schematic band structure of 1D modes in TI nanowires of smaller perimeter (L $\sim$ 100 nm) with two quantum flux value (0 and $\pm$\textit{h/}2\textit{e}). Fermi level (E$_F$) is only crossing a few subbands, and easily tunable across the Dirac point. The large 1D gap ($\Delta$$_{1D}$ = 17 meV) enables detection of 1D helical mode (red and blue) by locating E$_F$ at the Dirac point without crossing spin-degenerated 1D bands (black). (\textbf{b}) Magnetoconductance of device B (10 nm height $\times$ 50 nm width) at V$_G$ = -25V (Top), V$_G$ = -45V (Bottom, near the Dirac point) under parallel magnetic field. (\textbf{c}) Resistance (R) vs gate voltage (V$_G$) graph from device C (10 nm height x 40 nm width), with $\Phi$ = 0 (Blue) and $\Phi$ = \textit{h/}2\textit{e} (Red). The resistance maximum indicates the Fermi level close to the Dirac point. (\textbf{d}) Conductance ($\Delta$G) plot as a function of magnetic flux ($\Phi$) and gate voltage (V$_G$) (device C). A strong conductance peak is detected at half quantum flux / near the Dirac point (V$_G$ $\sim$-22V). Background conductance of G$_0$ (V$_G$) given V$_G$ is subtracted: $\Delta$G ($\Phi$, V$_G$) = G ($\Phi$, V$_G$) - G$_0$ (V$_G$). }
\end{figure}

As mentioned previously, TI nanowires of very small cross section (10 nm height $\times$ 50 nm width) is an ideal device to study helical transport near the Dirac point, due to their large 1D band gap (17 meV). The large 1D band gap also implies that only a few subbands exist below the initial Fermi level (Fig. 3a), facile to reach the Dirac point by gating. In fact, we can gate the devices to the Dirac point with relatively small gate bias (-20V - -40V), by finding resistance maxima in gate voltage sweep. From the magnetoconductance near the Dirac point, we find two strong conductance peaks at $\Phi$ = $\pm$\textit{h/}2\textit{e} (Fig. 3b), which is predicted as the 1D helical mode of gapless bandstructure (Fig. 1c). The different nature of gapped 1D mode ($\Phi$ = 0) and gapless 1D mode ($\Phi$ = $\pm$\textit{h/}2\textit{e}) can be clearly demonstrated by gate voltage sweep (Fig. 3c) and magnetic field / gate voltage sweep (Fig. 3d). At zero magnetic flux, resistance starts to diverge near the Dirac point (V$_G$ = -22V). However, at $\Phi$ = $\pm$\textit{h/}2\textit{e}, resistance converges while the Fermi level is tuned across the Dirac point (Fig. 3c). Indeed, the detailed scan of both magnetic flux ($\Phi$) and the gate voltage (Fig. 3d) reveals that there is a strong conductance peak when the Fermi level reaches at the Dirac point under $\Phi$ = $\pm$\textit{h/}2\textit{e}. The existence of 1D gapless electronic states under $\Phi$ = $\pm$\textit{h/}2\textit{e} directly proves the spin Berry's phase of the surface electrons, an unambiguous transport evidence of helical nature in the surface electrons\cite{Bardarson2010, Zhang2010PRL}.

\begin{figure*}
 \includegraphics[width=0.75\linewidth]{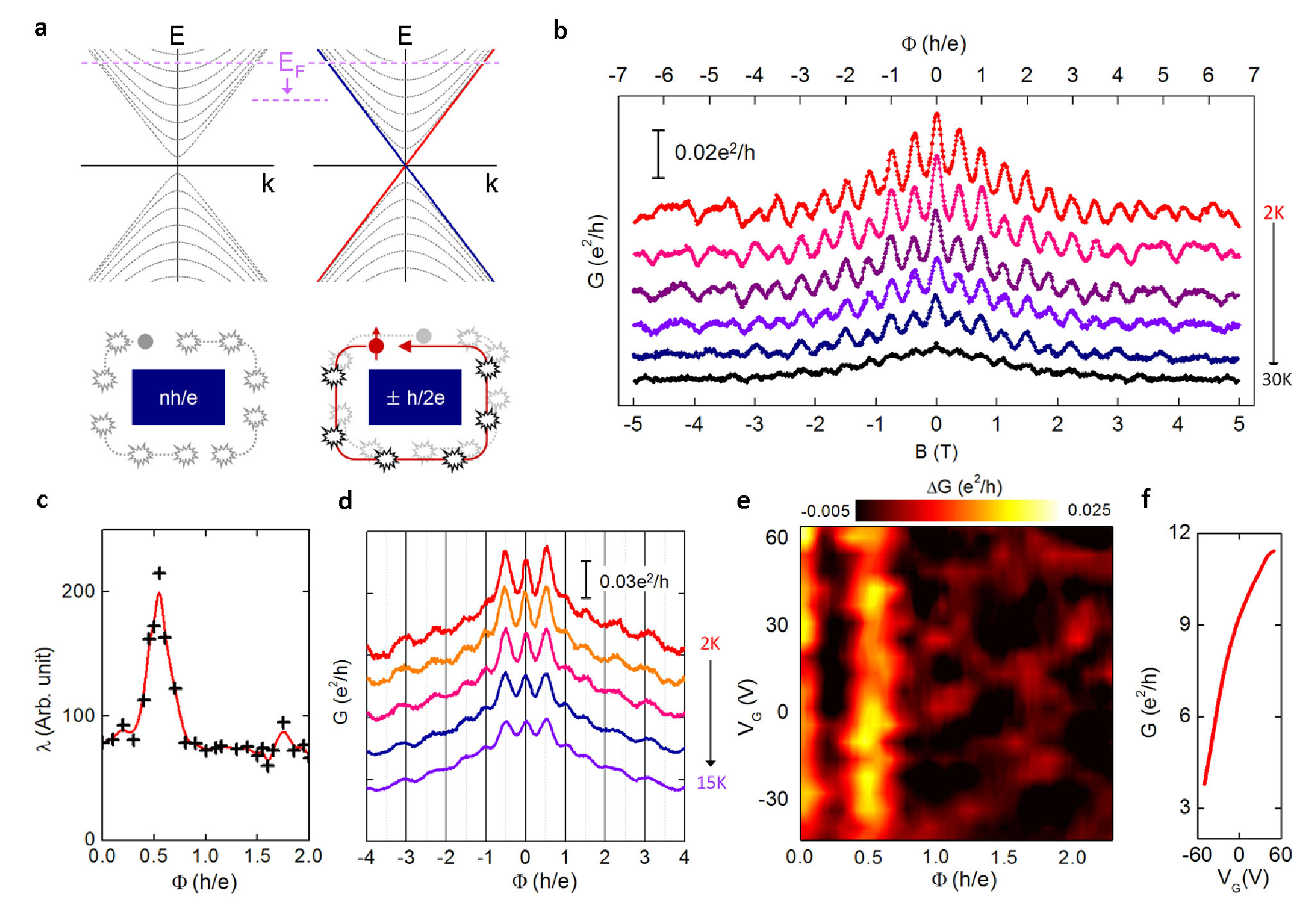}
 \caption{Magnetoconductance from 1D modes of surface electrons in disordered nanowires (\textbf{a}) Schematic band structure of 1D modes in disordered topological insulator nanowire. Addition of disorder localizes topologically trivial (gapped) 1D bands but helical 1D bands (red and blue) are not localized by strong disorder. Bottom left. Spin-degenerated 1D modes are localized due to strong disorder (gray). Bottom right, The helical 1D mode (red) at half quantum flux (\textit{h/}2\textit{e}) is protected from localization. (\textbf{b}) Representative quantum interferences of \textit{h/}2\textit{e} period oscillations (device D) in a parallel field, at different temperatures (2K$\sim$30K). (\textbf{c}) Numerical simulation of the localization length ($\lambda$) in disordered nanowires, with disorder strength \textit{W}=2$\Delta$ ($\Delta$ is the bulk energy gap). Detail of the simulation method is in reference\cite{Zhang2010PRL}. (\textbf{d}) A representative quantum interference of a disordered nanowire device (device E) in parallel field, at different temperatures (2K$\sim$15K). Two strong conductance peaks ($\sim$0.04\textit{e}$^2$\textit{/h}, total conductance = 3.3\textit{e}$^2$\textit{/h}) are observed at $\Phi$ = $\pm$\textit{h/}2\textit{e}, in addition to the suppressed oscillations ($<$0.01\textit{e}$^2$\textit{/h}) and the zero flux peak from WAL of bulk carriers. (\textbf{e}) Conductance ($\Delta$G) plot from a disordered nanowire device, as a function of magnetic flux ($\Phi$) and gate voltage (V$_G$) (device F). The peak at $\Phi$ = \textit{h/}2\textit{e} persists wide range of Fermi level, in parallel with WAL peak ($\Phi$ = 0) and suppressed oscillations. The peak height is 0.2$\sim$0.5$\%$ of the total conductance (\textbf{f}) Conductance change by back gating voltage (V$_G$) (device F).}
 \label{fig4}
\end{figure*}

We add comments on the conductance peak height of helical 1D mode, which can provide more insight of the topological 1D transport. In general, when the sample is in ballistic limit (along the length), quantized conductance is expected from the helical 1D states. In the previous example of HgTe/CdTe system, quantized conductance (2\textit{e}$^2$\textit{/h}) was observed when the channel length is comparable to the electron mean free path ($\sim$1 $\mu$m). A fraction of quantum conductance (0.3\textit{e}$^2$\textit{/h}) was observed from longer devices, implying the 1D conductance quantization is still depending on the electronic mobility in spite of the helical nature. In our case, electronic mobility of Bi$_2$Se$_3$ core-shell nanowires (~10$^4$ cm$^2$/VS) is less than that of HgTe system (10$^5$ cm$^2$/VS). The given mobility of our samples anticipates a fraction of quantum conductance (\textit{e}$^2$\textit{/h}) from 1D helical mode in Bi$_2$Se$_3$ nanowires when the device length is longer than the mean free path ($\sim$300 nm), which agrees with our experimental observation of 0.15\textit{e}$^2$\textit{/h} from 3 $\mu$m long device. 

In addition to the helical nature of surface electrons proven by 1D band structures, nanowire AB interferometers can provide a novel route to probe topological protection, even in the case when the device is no longer ballistic along the nanowire length. This is due to the coexistence of topologically trivial 1D modes (gapped) and helical 1D mode (gapless), which will be localized by disorders in different manners (Fig. 4a). In a clean nanowire (ballistic along nanowire perimeter), both gapped 1D modes and gapless 1D modes contributes to the AB oscillations, giving \textit{\textit{h/e}} period oscillations as discussed in Fig. 2. As the nanowire gets disordered, the ballistic condition along the nanowire perimeter is not held anymore thus \textit{\textit{h/e}} period oscillations are suppressed. However, AAS (\textit{h/}2\textit{e} period) oscillations rely on phase-coherent condition, existing in weakly disordered nanowires\cite{Bardarson2010, Zhang2010PRL} (Fig. 4b), similar to conventional metal cylinders\cite{Aronov1987}. The interesting part is strongly disordered regime, where scattering by impurities is strong enough to suppress \textit{h/}2\textit{e} period oscillations. All the quantum interferences from ordinary (spin-degenerated) electronic states will diminish by strong disorder. However, helical 1D states (with no spin degeneracy) will not be localized due to the topological protection\cite{Bardarson2010, Zhang2010PRL} (Fig. 4a), resulting in positive peaks in magneto-conductance. One important premise here is that such disorder should not create inelastic scattering events. From the TI nanowires without a protective layer, we only observe the conductance peak at zero flux corresponding to weak antilocalization of bulk carriers. We believe that the chemically modified Bi$_2$Se$_3$ surface introduces many inelastic impurities destroying TRS in the surface states. To induce strong disorder without affecting the Bi$_2$Se$_3$ surface, we kept core-shell nanowire devices under inert atmosphere where aging at room temperature would create more disorder in the TI nanowire core but the surface would remain chemically stable under the Se shell.

The numerical simulation\cite{Zhang2010PRL} of a disordered TI nanowire (Fig. 4c) provides a physical picture based on localization by disorders. With strong disorder (Disorder strength \textit{W}=2$\Delta$, $\Delta$ is the bulk energy gap) enough to localize 1D mode at most magnetic flux values, localization length peaks at $\Phi$ = \textit{h/}2\textit{e}, anticipating two conductance peaks peaking at $\Phi$ = $\pm$\textit{h/}2\textit{e}\cite{Zhang2010PRL}. We note that in simulation, there is no conductance peak anticipated at higher orders of half quantum flux (i.e. 3\textit{h/}2\textit{e}). The absence of high order half quantum flux peaks is described as a local TRS breaking effect near impurities under a large parallel magnetic field.

In transport experiments of a disordered TI nanowire device (introduced by sample aging), we observe two strong conductance peaks at $\Phi$ = $\pm$\textit{h/}2\textit{e} clearly distinguishable from suppressed oscillations (Fig. 4d), as predicted by the simulation (Fig. 4c). The peaks at $\Phi$ = $\pm$\textit{h/}2\textit{e} are even larger than the zero-field peak (containing bulk WAL effect) and, thus unlikely the remains of \textit{h/}2\textit{e} period oscillations. In addition, we study gate-dependence of the conductance peaks. In the strong disorder limit, the gapped modes are localized by disorders. However, the gapless 1D helical mode still contributes to the quantum interference while the quantum interference from the gapped modes would be averaged out. Therefore, we expect to observe peaks at $\Phi$ = $\pm$\textit{h/}2\textit{e} regardless of the location of the Fermi level. In our experiments, the conductance peaks observed in multiple disordered nanowire devices always exist at $\Phi$ = $\pm$\textit{h/}2\textit{e} (Fig. 4e) in spite of large Fermi level change by gating (Fig. 4f). This is naturally anticipated from the gapless bandstructure of helical 1D mode\cite{Bardarson2010, Zhang2010PRL}. This concludes that the helical 1D mode of no spin degeneracy is robust against disorder strong enough suppress all the other quantum interference, a clear demonstration of topologically protected transport of spin-textured surface electrons.

\begin{figure}
\centering
\includegraphics[width=88mm]{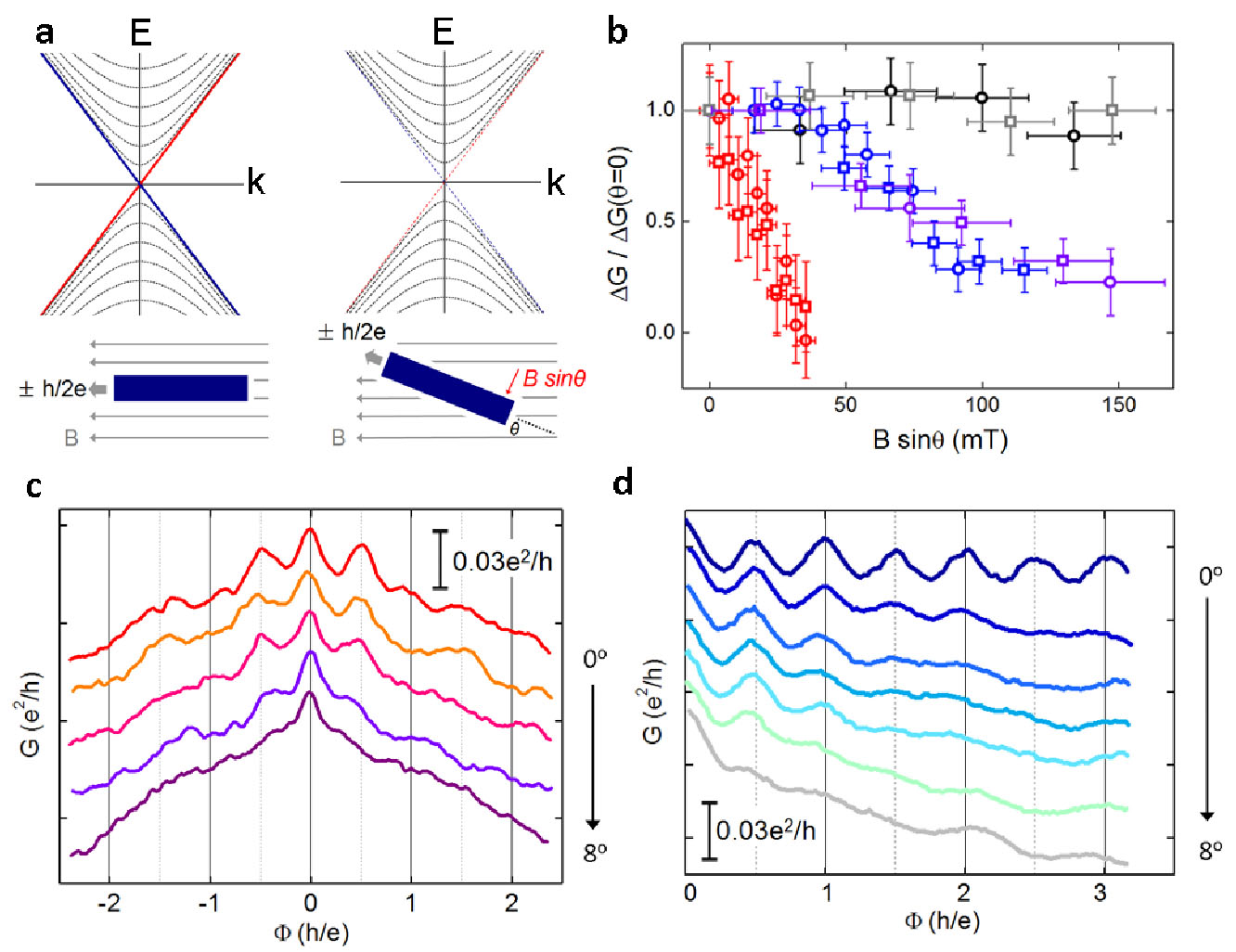}

\caption{Breaking time reversal symmetry by a perpendicular magnetic field (\textbf{a}) Left: schematic band diagram of a disordered TI nanowire at $\Phi$=$\pm$\textit{h/}2\textit{e}. Under strong disorder, helical 1D modes (red and blue) are not suppressed. Right: Addition of perpendicular magnetic field  (B sin$\theta$), by device rotation breaks TRS condition at $\Phi$=$\pm$\textit{h/}2\textit{e}, resulting in localized helical 1D modes in a disordered TI nanowire. (\textbf{b}) Quantum interference peak ($\Delta$G) evolution by perpendicular magnetic field. \textit{\textit{h/e}} period oscillations from clean nanowire devices correspond to black (device A) and gray (device G) data points, while \textit{h/}2\textit{e} period oscillations from clean nanowires correspond to blue (device A) and purple (device G) data points. Helical mode at $\Phi$=$\pm$\textit{h/}2\textit{e} (device F) corresponds to red data points. (\textbf{c}) Quantum interferences (helical 1D peaks at $\Phi$=$\pm$\textit{h/}2\textit{e}) at different angles from a disordered nanowire (device F). (\textbf{d}) \textit{h/}2\textit{e} period oscillations at different angles (device A), revealing the peak suppression is not angle-dependent but dependent on absolute-field.}
\end{figure}

Finally, the 1D helical mode peaks are tested under TRS breaking condition, by a perpendicular magnetic field\cite{Konig2007}. The nanowire devices, initially aligned along the magnetic field direction (Fig. 5a left), are tilted to create a small perpendicular field (Fig. 5a right). The magnetic field perpendicular to the nanowire length is breaking TRS, and is expected to localize the helical mode. We first monitor angular dependent oscillations from clean nanowire devices as a control example (Fig. 5b). \textit{h/e} period oscillations from 1D DOS effect do not change much by the perpendicular field within a few hundreds millitesla (Fig. 5b). This is consistent with the theoretical explanation that the \textit{h/e} period oscillations are not from interference but from the 1D bandstructure which is insensitive to the TRS breaking. In contrast, the \textit{h/}2\textit{e} oscillations are from the interference between electrons along two different time-reversed paths, therefore these oscillations would be affected by TRS breaking to some degree. The experimental data show that the amplitude of \textit{h/}2\textit{e} oscillations decreases as the perpendicular field increases ($\sim$100 mT) (Fig. 5b,d). The most fragile feature observed under the TRS breaking condition is the helical 1D mode peaks at the first half quantum flux ($\Phi$ = $\pm$\textit{h/}2\textit{e}) persisting in the disordered limit. They disappear at the perpendicular magnetic field of $\sim$20 mT (Fig. 5b-c). The fragileness against TRS breaking provides another evidence of the topological nature of the helical 1D mode, which is very similar to the magneto-conductance of the HgTe quantum spin Hall edge state\cite{Konig2007}.

TI nanowire AB interferometer demonstrates a quantum device of the helical nature and the topologically distinctive transport property (robustness against disorder). A 1D “coherent electron waveguide” of surface electrons, tailored by magnetic flux quantum, provides unique electronic states crucial to realize topological electronics in nanowires. For example, the nanowire interferometer with helical 1D mode is expected to host 1D topological superconducting states immune to disorder\cite{Cook2011}, serving as an attractive candidate to manipulate Majorana zero-modes in 1D nanostructures\cite{Alicea2012, Franz2013}. Appropriate material design was critical to study the exotic properties of the topological surface states. The clean TI surface protected by an in-situ Se coating allows us to explore high mobility transport with coherent topological nature. This core-shell heterostucture nanowire device offers an ideal candidate for a “topological quantum device,” making use of the exotic nature of the surface electrons.

\subsection{Acknowledgments}

\begin{acknowledgments}
We gratefully acknowledge discussions with D. Goldhaber-Gordon, P. Kim, S.-B. Chung, J. R. Williams, K. Lai, and S. C. Zhang, and thank K. J. Koski for the help in the manuscript preparation. Y.C. acknowledges the supports from the Keck Foundation and DARPA MESO project (No. N66001-11-1-4105). X.L.Q acknowledges the supports from DARPA MESO project (No. N66001-11-1-4105).
\end{acknowledgments}

\end{document}